\documentclass[a4paper,12pt]{article}
\usepackage{graphicx}
\usepackage{psfrag}
\def\journal#1#2#3#4{{\it #1} {\bf #2} (#3) #4}
\def\epj{Euro. Phys. Jour.}
\def\prl{Phys. Rev. Lett.}
\def\pl{Phys. Lett.}
\def\np{Nucl. Phys.}

\def\pr{Phys. Rev.}
\def\prp{Phys. Rep.}
\def\nc{Nuovo Cim.}

\def\le{\ell}
\def\ups{\Upsilon}
\def\tcups{t \rightarrow c \, \ups}
\def\tbw{t \rightarrow b \, W}
\def\bb{\bar{b}}
\def\cb{\bar{c}}

\def\ub{\bar{u}}
\def\m{{\cal M}}

\def\b{{\cal X}}
\def\br{{\it B}}
\def\o{{\cal O}}

\def\ot{\tilde{\cal O}}
\def\c{{\cal C}}
\def\f{{\it F}}
\def\fl{{{\cal F}^\prime}}
\def\l{{\cal L}}
\def\t{{\cal T}}

\def\g{\gamma}
\def\gl{\Gamma}
\def\gh{\hat{G}}
\def\mh{\hat{m}}
\def\mch{\mh_c}
\def\mbh{\mh_\b}
\def\mbqh{\mh_b}
\def\mwh{\mh_W}
\def\muh{\mh_\ups}

\def\gbh{\hat{\gl}_\b}

\def\la{\lambda}
\def\lp{{\lambda^\prime}}
\def\lpijk{{\lambda_{ijk}^\prime}}

\def\lpp{{\lambda^{\prime \prime}}}
\def\lppijk{{\lambda_{ijk}^{\prime \prime}}}
\def\gf{\hat{G}_F}

\def\tf{{\tilde{f}}}
\def\td{{\tilde{d}}}
\def\tl{{\tilde{\le}}}
\def\mrp{{$-R_{\rm p}$}}
\def\prp{{$+R_{\rm p}$}}

\title{$\tcups$ Within and Beyond the Standard Model\\~}
\author{L. T. Handoko$^{1,2}$ \footnote{Alexander von Humboldt Fellow. 
E-mail: handoko@mail.desy.de, handoko@p3ft.lipi.go.id}
\hspace{2mm} and \hspace{2mm} Cong-Feng Qiao$^{3,4}$
\footnote{Alexander von Humboldt Fellow.
E-mail: qiaocf@mail.desy.de, qiaocf@ccastb.ccast.ac.cn}\\\\
{\small $^1$Theorie Gruppe, DESY, Notkestr. 85, D-22607 Hamburg, Germany}
\\ [6pt]
{\small $^2$Lab. for Theor. Physics and Mathematics, LIPI Kom. 
Puspitek Serpong P3FT--LIPI,}\\
{\small Tangerang 15310, Indonesia}
\\[6pt]
{\small $^3$II Institut f\"ur Theoretische Physik, Universit\"at
Hamburg, Luruper Chaussee 149,}\\
{\small D-22761 Hamburg, Germany}
\\[6pt]
{\small $^4$CCAST (World Laboratory), P.O. Box 8730, Beijing 100080, 
P. R. China}}
\date{}
\begin{document}
\maketitle
\thispagestyle{empty}
\begin{abstract}

The top quark decay process $\tcups$ is studied in the Standard Model
and the Minimal Supersymmetric Standard Model with conserved and
violated $R-$parity. In dealing with the heavy quarkonium, $\ups$,  
production, we take both color-singlet and -octet mechanisms 
into consideration. With numerical analysis it is found that within the
Standard Model the decay rate of the concerned process is beyond the border
of reachable level of the next round of experiment, though it is much larger
than that of the rare top decays, while extra new contributions from
beyond the Standard Model could enhance the decay rate by several orders.
It is also noticed that a precise measurement of the charged Higgs mass
without $\tan\beta$ dependence is possible via this process.
\vspace*{7mm}

\noindent
PACS number(s) : 14.65.Ha, 12.60.-i
\end{abstract}

\clearpage
\section{Introduction}

The discovery of the top quark in 1995 \cite{r1,r2,r3} marked the triumph
of the Standard Model(SM) of particle physics. Since the discovery, more
interest on researches in top quark physics is stimulated in both
experimental and theoretical respects. Because of its large mass which is
close to the electroweak scale, the top quark physics research may play an
important role in the study of the electroweak symmetry breaking and
therefore of the origin of the fermion masses. Through the study on it,
especially its rare decays, people also hope to find clues of possible
"new physics".

Within the framework of SM the dominant top quark decay mode is $t
\rightarrow b W$, which has a time scale $\tau_W$ much shorter than that
of the non-perturbative QCD effect. This character of top quark makes it
behaves almost like a free particle, which is helpful to the precise study
of it. With the operation of CERN Large Hadron Collider (LHC) in the
future there will be enormous top quark events obtained every year and
that provides it practically possible to make a more detailed study on
many properties of top quark.

Up to now, the rare top decays have been investigated by several groups
within the SM\cite{2hdm, mele} and beyond, e.g. in 2-Higgs Doublet
Model(2HDM) \cite{2hdm, gunion} and Supersymmetric(SUSY) models
\cite{susy}. In SM, the branching ratios are heavily suppressed because of
the Glashow-Iliopoulos-Maiani(GIM), typically ${\it B}(t \rightarrow c \, H) 
< {10}^{-13}$ \cite{mele}. Though they may be greatly enhanced by new 
contributions from
beyond the SM, the rare top decays are still quite small with only few
cases up to the measurable levels of the near future top producing
machines, like the upgraded Tevatron and LHC \cite{fermilab}. The decay
channel $\tcups$ was once investigated for the aim of studying
CP-violation and it was also claimed having an accessible rate for the
future experiments based on a quite rough assumption on the
non-perturbative sector of the heavy quarkonium \cite{lth}, which leads to
about two orders of magnitude overestimate of the rate, to be seen in the
following.

In this paper, we will investigate the top decay $\tcups$ process with
taking into account both the color-singlet and -octet mechanisms in the
$\ups$ production calculation.  We find the process is worth being
carefully investigated in testing the SM and especially in searching new
physics signatures for the following arguments. First, comparing with the
meson decays, the top quark decays possess less theoretical uncertainties,
e.g. without non-perturbative long distance influence because of its large
mass scale. Second, the concerned decay mode occurs at tree-level, which
may cause the decay rate larger than that of rare top decays. Third,
theoretically the heavy quarkonium, $\Upsilon$, can be treated in a
relatively high precision by virtue of the Non-relativistic QCD(NRQCD)
factorization scheme \cite{qcd}. Besides, from the experimental point of
view, heavy quarkonia have clean signatures in their leptonic decays.

The paper is organized as follows. In section 2 we present the formalism
of the concerned process $\tcups$ in a general framework. In section 3 the
obtained formalism is applied to some specific models, the SM and Minimal
SUSY Standard Model (MSSM) with and without $R-$parity violation (MSSM$\mp
R_{\rm p}$). The numerical evaluation is proceeded in section 4, and in
the last section some conclusions are made.

\section{Formalism}

To lowest order, it is quite clear that the interested decay channel can
be induced by intermediate bosons, for example the $W$ in the SM. In order
to be more general and convenient to incorporate contributions from
different models, in the following we give out the expression of the decay
amplitude in a universal form.

The tree-level amplitude
of $\tcups$ could be expressed as 
\begin{eqnarray}
\label{11}
  \m & = & 2 \sqrt{2} G_F \, \mwh^2 \, (V_{cb}^\ast V_{tb})
  \sum_\b \, 
  \left< \ups + R \left| \sum_{i=\pm} \f_{i\b} \c_i \, \o_i^\b \right| t 
\right> \; ,
  \label{eq:m}
\end{eqnarray}
where $\b$ stands for the intermediate particles ($W-$boson, charged Higgs,
etc.), which are normalized by the standard $W$-boson coupling for the
sake of convenience; $R$ represents the remnant in the final states in
respect to the $\Upsilon$ production. For abbreviation, the caret means
the corresponding quantity is normalized by the top quark mass $m_t$, i.e.
$\mwh = {m_W}/{m_t}$. The four-quark non-local operators appearing in Eq.
(\ref{eq:m}) are of the types 
\begin{eqnarray}
\o_+^\b & = & 
\left[ \cb \, \left( \sum_{i} g_i G_i \right) b \right] 
\left[ \bb \, \left( \sum_{i} {g^\prime_i} G_i \right) t \right] \; , 
\label{eq:op} \\[6pt]
\o_-^\b &=&
\left[ \cb \, \left(\sum_{i} h_i H_i \right) t \right] 
\left[ \bb \, \left(\sum_{i} {h^\prime_i} H_i \right) b\right]
\; ,
\label{neutral}
\end{eqnarray}
where ${g_\pm^{(\prime)}}$ and ${h_\pm^{(\prime)}}$ represent the 
couplings of the fermion interactions, the $G_i$ and $H_i$ are the 
possible tensor structures of the currents. It is noticed that 
although in principle the quarkonium can be generated through 
the neutral current of form (\ref{neutral}), however, it is obvious 
that this kind of possible operators, which produce $\ups$ from a 
neutral heavy vector boson, are of kinematically disfavored
ones in leading order since the possible large virtuality of intermediate
particles. Therefore, it is proper we restrict ourselves to the
dominant contribution, i.e. form (\ref{eq:op}). As well, in our
discussion the penguin process of the quarkonium production via gluon
fragmentation is also dropped because that is suppressed by both the GIM
mechanism and the smallness of the higher order quarkonium Fock states
\cite{qiao}.

In the top quark rest frame, the $\f_{i\b}$, which denotes the
propagator of $\b$, can be expressed as $\f_{i\b}
= \f_i (1,\mch,\muh,\mbh,\gbh)$ with
\begin{eqnarray}
  \f_+(1,x,y,z,w) & \equiv & 2 \, \left( 
    1 + x^2 - \frac{1}{2} y^2 - 2 z^2 + 2 i \, z \, w \right)^{-1} \; , \\
   \f_-(1,x,y,z,w) & \equiv & 
    \left( y^2 - z^2 + i \, z \, w \right)^{-1} \; .
  \label{eq:f}
\end{eqnarray}
This is obtained by performing the integrations over the final state phase
space at the level of decay width, while adding the unstable particle 
decay width to the propagator, and, here and in the after, the lowest 
order approximation relations in the quarkonium system, $E_\ups = 2
E_b$ and $m_\ups = 2 m_b$ are taken, or in other words, in the effective
Lagrangian method the relative movement effects of heavy quarks within the
quarkonium appears only in the nonperturbative sector. 
Since we only consider the $\o_+^\b$ in further discussion, we can
simply drop the subscript $i$ in $\f_{i\b}$, and then $\f_\b$ will 
represent $\f_{+\b}$ for short.

With performing the Fierz transformation on (\ref{eq:op}) and considering
of only the suitable operators in producing $\Upsilon$, the operator
$\o_+^\b$ could be read as
\begin{equation}
  \o_+^\b = \frac{1}{N_c} \bar{\o}_-^\b + \frac{1}{2} \ot_-^\b \; , 
\end{equation}
where $N_c$ is the number of colors and
\begin{eqnarray}
\bar{\o}_-^\b & = &
\sum_{i=\pm} g_i^\b {g^\prime_i}^\b \, 
\left[ \cb \, \g_\mu P_i \, t \right] 
\left[ \bb \, \g^\mu P_i \, b \right] \; , 
\label{eq:om} \\[6 pt]
\ot_-^\b & =& 
\sum_{i=\pm} g_i^\b {g^\prime_i}^\b \, 
\left[ \cb \, \g_\mu P_i \, \la^a \, t \right] 
\left[ \bb \, \g^\mu P_i \, \la^a \, b \right] \; ,
\label{eq:ot} 
\end{eqnarray}
with $\la^a$ representing the $SU(3)$ Gell-Mann matrices. Here 
$P_\pm \equiv {(1 \pm \g_5)}/2$ are chirality projection operators.
At the first sight Eq.(\ref{eq:om}) appears to be the same structure with 
(\ref{neutral}), actually they possess different natures in describing 
quarkonium production. Therefore, the operators in Eq.(\ref{11}) may
be rearranged as
\begin{equation}
  \sum_{i=\pm} \c_i \, \o_i^\b = 
  \left( \c_-  {\o}_-^\b \, + \frac{1}{N_c} \c_+  \bar{\o}_-^\b \right) + 
  \frac{1}{2} \c_+ \, \ot_-^\b \; ,
\end{equation}
where the first term is called the color suppressed one. Next, by
taking only the leading-order(LO) electroweak effects 
into account one could replace $\c_\pm$ by
\begin{equation}
  \c_+ = 1 \; \; , \; \;  \c_- = 0 \; .
  \label{eq:cpm}
\end{equation}
The higher order corrections to these values, either from the QCD or
higher order electroweak corrections, should be small because of the large
top mass scale. Whereas, together with the arguments in before, the $\c_-$
term can definitely be discarded without losing accuracy.

\begin{figure}[t]
\centering
\includegraphics[scale=0.5]{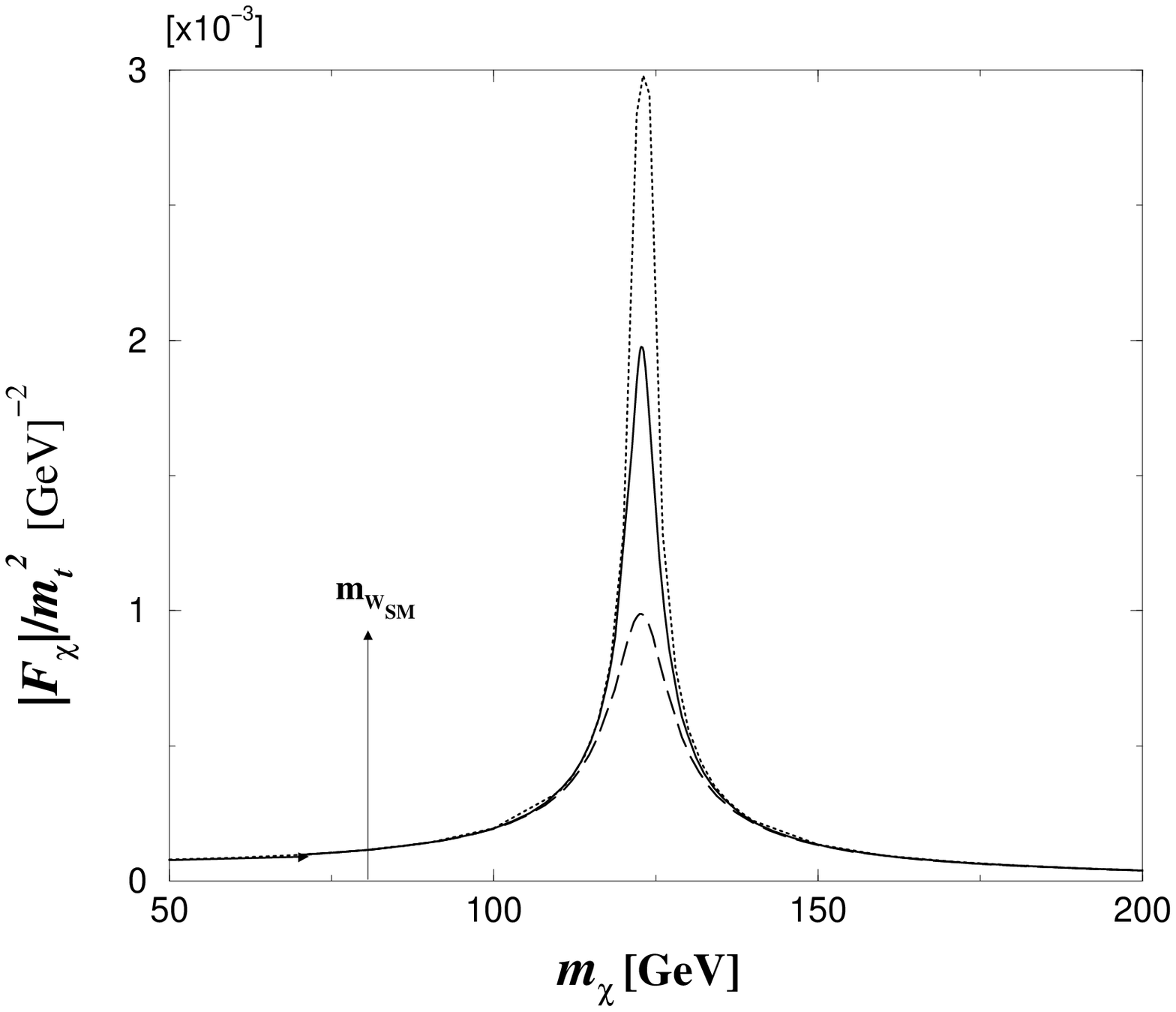}
\caption{
$|F_\chi|/m_t^2$ as a function of $m_\chi$ with $\Gamma_\chi
= 2\Gamma_{W_{SM}}$ (solid line), $\Gamma_\chi = 4 \Gamma_{W{SM}}$
(dashed line), and $\Gamma_\chi = \Gamma_{W_{SM}}$ (dotted line). The
arrow indicates the value of $m_{W{SM}}^{-2}$.}
 \label{fig:afb}
\end{figure}

In the NRQCD framework of factorization, the heavy quarkonium production
can be properly separated into two sectors, the perturbative QCD
calculable part and the non-perturbative universal matrix elements. The
latter is ordered in the relative velocity, $v$, of the quarks within the
quarkonium system. According to the NRQCD, a generic $S-$wave $\ups$ state
can be schematically described by the Fock state decomposition \cite{qcd}
\begin{eqnarray}
  \left| \ups \right> & = & 
  {\cal O}(1) \left| b\bb [^3S_1^{(1)}] \right> + 
  {\cal O}(v) \left| b\bb [^3P_J^{(8)}] \, g \right> + 
  {\cal O}(v^2) \left| b\bb [^1S_0^{(8)}] \, g \right>
  \nonumber \\
  & & +
  {\cal O}(v^2) \left| b\bb [^3S_1^{(1,8)}] \, g g \right> +
  {\cal O}(v^2) \left| b\bb [^3D_J^{(1,8)}] \, g g \right> +
  \cdots \; ,
  \label{eq:fock}
\end{eqnarray}
where the usual spectroscopic notation $^{2S+1}L_J$ is used, and the
color states are labeled by (1) and (8) superscripts for color-singlet
and -octet, respectively. It should be noted that in order to express 
the production rate in terms of NRQCD matrix elements rather than the 
quarkonioum wavefunction, the $b$ and $\bb$ must be created with a 
separation small compared to the size of the wavefunction, which
is of order $1/{m_b v}$. This condition can be expressed as
\begin{equation}
  \frac{1}{m_t^2} | \f_\b | \ll \frac{1}{(m_b v)^2} \; ,
  \label{condition}
\end{equation}
as is shown in Fig. \ref{fig:afb}. Taking the typical value of $v^2
\approx 0.08$, the value of the RHS of eq.(\ref{condition}) is
about 0.5 $\rm{GeV}^{-2}$ and it is obvious, from the figure, 
that the entire region of $m_\b$ is allowed for the presumed 
magnitude of $\b$'s decay width, i.e. $\Gamma_\b \sim$ ${\cal{O}}(1)$GeV.

Based upon preceding preparations the $\tcups$ amplitude can now be 
expressed as  
\begin{eqnarray}
  \m & = & \sqrt{2} G_F \, \mwh^2 \, (V_{cb}^\ast V_{tb})
  \sum_\b \, \sum_{i=\pm} \f_\b \, g_i^\b {g^\prime_i}^\b \c_+  
  \nonumber \\
&& \times\left\{\frac{1}{2} \, \left[ \cb \, \g_\mu P_i \, \la^a \,t
\right] \,  (\t_8^\mu)^a + \frac{1}{N_c} \, \left[ \cb \, \g_\mu P_i \, t
\right] \, \t_1^\mu \right\} \; 
  \label{eq:ml}
\end{eqnarray}
with $\t_1$ and $\t_8^a$ stand for the color-singlet and -octet
non-perturbative matrix elements, 
\begin{eqnarray}
\t_1^\mu & \equiv & 
\left< 0 \left| \bb \, \g^\mu \, b \right| \ups + X \right> \; ,
\label{eq:t1} \\ 
(\t_8^\mu)^a & \equiv & 
\left< 0 \left|\bb \, \g^\mu\la^a \, b \right|\ups + X \right>\; .
\label{eq:t8} 
\end{eqnarray}
Here, $X$ represents the possible soft gluon radiations in the process
of forming the Onium.

According to expansion (\ref{eq:fock}), the contribution of higher order 
Fock states evolving into the quarkonium by soft gluon(s) emission or 
absorption are superficially suppressed by some orders of the small
quantity of $v^2$ relative to the leading singlet one, however, they may 
play an important role in special cases \cite{surplus}, where the higher 
Fock state involved processes are compensated by the enhancements
originated from other kinematical variables or coupling constant. Notice 
that the quarkonium production process interested in here may have a 
factorized form provided by the NRQCD, in the following for the sake
of later convenience the matrix elements in eqs. (\ref{eq:t1}) and 
(\ref{eq:t8}) are squared with the soft radiations being summed over 
and then parameterized as
\begin{eqnarray}
\sum_X (\t_1^\mu)^*  \t_1^\nu & \equiv & m_\ups^2 \, 
(f^{1}_\ups)^2 \, \epsilon_\ups^{* \mu} \, \epsilon_\ups^\nu\; ,
  \label{eq:part1} 
\end{eqnarray}
\begin{eqnarray}
\sum_X (\t_8^{\mu a})^* \t_8^{\nu b} 
\equiv \delta^{ab} \left[
(f^{8,^1S_0}_\ups)^2 p_\ups^\mu \, p_\ups^\nu +
m_\ups^2\, (f^{8,^3S_1}_\ups)^2 \, \epsilon_\ups^{* \mu} \, 
\epsilon_\ups^\nu\; + m_\ups^2\, (f^{8,^3P_1}_\ups)^2 \,
\epsilon_\ups^{* \mu} \, \epsilon_\ups^\nu \right] \; . ~
  \label{eq:part8} 
\end{eqnarray}
Here, the constants $f^{1; 8, ^{2S+1}L_J}_\ups$, the ``decay constants'', 
stand for the 
non-perturbative parts in the quarkonium producton with a mass dimension,  
which can be related to the vacuum-to-vacuum matrix elements of the 
NRQCD 4-fermion operators as shown later; the $p_\ups$ and 
$\epsilon_\ups^\mu$ are the momentum and polarization vector of $\ups$. 

As compared with the color-octet matrix elements, which may now only be
determined through experiment, the color-singlet part can be determined by
several means with more accuracy, e.g. from the quarkonium leptonic decay
$\ups \rightarrow \le^+ \, \le^-$, etc. Up to the next to leading
order calculation, the decay width is \cite{schuler}
\begin{equation}
  \gl(\ups \rightarrow \le^+ \, \le^-) =
  \frac{4 \pi}{3} \, Q_b^2 \, \alpha(m_\ups)^2 \, 
  \frac{(f^1_\ups)^2}{m_\ups} 
  \la\left(1,\frac{m_\le}{m_\ups}\right)^{1/2}
  \left( 1 + \frac{2 m_\le^2}{m_\ups^2}  \right)
  \left( 1 + \frac{16 \alpha_s(m_\ups)}{3 \pi} \right) \; ,
\end{equation}
where $\la(1,x) \equiv \la(1,x,x)$ is the triangle function and 
$\la(1,x,y) = 1 + x^4 + y^4 - 2 (x^2 + y^2 + x^2 y^2)$.  With experimental
values \cite{pdg} as input, using the running couplings $\alpha(m_\ups) =
1/{133}$, $\alpha_s(m_\ups) = 0.178$ and $m_\ups = 9460.37 \pm 0.21$
(MeV), one immediately obtains the average value of $f^1_\ups$ 
\begin{equation}
  \left(f^1_\ups \right)^2_{\rm av.} = 0.364^{+2.3\times 10^{-4}}_
  {-2.6\times 10^{-4}} \; {\rm GeV}^2 \; .
  \label{eq:f1}
\end{equation}
Here, the uncertainty mainly comes from $\gl_{\rm exp.} (\ups \rightarrow
\le^+ \, \le^-)$.

The color-octet decay constants defined in Eq. (\ref{eq:part8}) are
correlated with the NRQCD matrix elements, which are phenomenologically
determined from experiments, whereas some quantitative diversities of
different fit still exist among them. Taking values fitted in 
Ref.\cite{qcd8}, we have
\begin{equation}
  (f_\ups^{8, ^3S_1})^2 =
{\left< 0\left|\o_\ups [^3S_1^{(8)}]\right|0\right>}/{(12 m_\ups)} 
 = (5.2 \pm 1.4) \times {10}^{-5} \, {\rm GeV}^2 \; ,
\end{equation}
\begin{equation}
  (f_\ups^{8, ^1S_0})^2 =
{\left< 0 \left| \o_\ups [^1S_0^{(8)}] \right| 0 \right>}/{4 m_\ups}
= (5.2 \pm 1.4) \times {10}^{-5} \, {\rm GeV}^2 \; ,
\end{equation}
\begin{equation} 
(f_\ups^{8, ^3P_1})^2 =
{2\left< 0\left|\o_\ups[^3P_1^{(8)}]\right| 0\right>}/{3 m_\ups^3}
= (3.2 \pm 2.8) \times {10}^{-4} \, {\rm GeV}^2 \;.
\label{oconst}
\end{equation}
Here, $\left< 0 \left| \o_\ups [X^{(8)}] \right| 0 \right>$ are the 
non-perturbative color-octet matrix elements defined in \cite{qcd}. 
In evaluating the $f_\ups^8$s the approximation of heavy quark spin
symmetry has been used, i.e.  
\begin{eqnarray}
  \left< 0 \left| \o_\ups [^3S_1^{(8)}] \right| 0 \right> 
  & \approx &
  3 \left< 0 \left| \o_\ups [^1S_0^{(8)}] \right| 0 \right> \; , \\ 
  \left< 0 \left| \o_\ups [^3P_J^{(8)}] \right| 0 \right> 
  & \approx & 
  (2 J + 1) \left< 0 \left| \o_\ups [^3P_0^{(8)}] \right| 0 \right> 
  \; .
\end{eqnarray}
It should be noted that the above calculations on the ``decay
constants'' is
just an order estimation, especially the color-octet part where large
uncertainties still survive. Therefore, the results derived in this paper
are correspondingly accurate up to about an order.

\section{Applying to Models}

With the formalism obtained in preceding section, it is straightforward to
make an estimation of the branch ratios of the process $\tcups$ in some
specific models. For top quark decays, because the overwhelming mode is
$\gl(\tbw)$, at least in standard model, the branching ratio
$\br(\tcups)$ can be expressed as
\begin{equation}
  \br(\tcups) = \frac{\gl(\tcups)}{\gl(\tbw)} \; ,
  \label{eq:br}
\end{equation}
with the assumption that $\br(\tbw) \approx 1$ in all cases in the
discussion.

To next-to-leading order in strong coupling constant, the decay width of
$\tbw$ is \cite{tbw}
\begin{equation}
  \gl(\tbw) = \frac{\gh_F \, m_t}{8 \sqrt{2} \pi} 
        \left| V_{tb} \right|^2 \la(1,\mbqh,\mwh)^{1/2} \, 
        \fl_W \, 
        \left[ 1 - \frac{2 \alpha_s(m_t)}{3 \pi} 
        \left( \frac{2 \pi^2}{3} - \frac{5}{2} \right) \right] \; ,
\end{equation}
where $\fl_W = \fl(1,\mbqh,\mwh)$ and
\begin{equation}
        \fl(1,x,y) \equiv (1 - x^2)^2 + (1 + x^2) y^2 - 2 y^4 \; .
        \label{eq:fp}
\end{equation}

Based upon Eq. (\ref{eq:ml}) and considering of the possible color-octet
contributions, the decay width of $\tcups$ has the form
\begin{equation}
  \gl(\tcups) = \frac{\gf^2 \, \mwh^4 m_t}{8 \pi} \, 
  \left| V_{cb}^\ast V_{tb} \right|^2 \,  
  \la(1,\mch,\muh)^{1/2} 
  \left[ \left( \frac{\hat{f}_\ups^1}{N_c} \right)^2 \, \Delta_1 
    + 8 \sum_{\ups^8} (\hat{f}_\ups^8)^2 \, \Delta_8 
  \right] \; .
  \label{eq:dwtcups}
\end{equation}
Here, $\Delta_j$ contain contributions from various $\b-$mediated
diagrams, and
\begin{equation}
    \Delta_j  = 
    \fl_j \, \sum_{i=\pm} \left| G_i \right|^2
    + a_j \mch \, \muh^2 \, {\rm Re} \left[ \prod_{i=\pm} 
      G_i \right] \;  
     \label{eq:delta}
\end{equation}
with
\begin{eqnarray}
  G_i & \equiv & \sum_\b g_i^\b \, {g^\prime_i}^\b \, \f_\b \; , \\
  (a_1, a_{8 [^3S_1]}, a_{8 [^1S_0]}, a_{8 [^3P_1]}) & = & 
   (-12, -12, 4, -12) \; . 
\end{eqnarray}
Here, the auxiliary functions $\fl_j$ are
\begin{eqnarray}
  & & \fl_1 = \fl_{8 [^3S_1]} = \fl_{8 [^3P_1]} = \fl(1,\mch,\muh) \; , 
  \label{eq:fl}
\end{eqnarray}
which in principle could be complex numbers, and
\begin{eqnarray}
  & & \fl_{8 [^1S_0]} = (1 - \mch^2)^2 - (1 + \mch^2) \muh^2 \; .
  \label{eq:fl2}   
\end{eqnarray}
In the following, the explicit form of the function $\Delta_j$ in concrete
models will be given.

\subsection{SM}

In the SM, to leading order the concerned decay process is generated  by
the $W-$boson mediated diagram via interaction
\begin{equation}
  \l_{W^\pm} = \frac{g}{\sqrt{2}} \, V_{ij} \, 
      \left[ \ub_i \, \g^\mu P_- \, d_j \right] \, W_\mu^+ + {\rm h.c.} \; , 
\end{equation}
where $V_{ij}$ denotes the CKM matrix element. The function $\Delta_j$
defined in the previous section thereof is
\begin{eqnarray}
    \Delta_j^{\rm SM} & = & \left| \f_W \right|^2 \, \fl_j \; 
    \label{eq:deltaw} 
\end{eqnarray}
with
\begin{eqnarray}
  G_+ = 0 \; , \;  
  G_- = \f_W  \; .
\end{eqnarray}
Here, the definitions of $\f_W$ and $\fl_j$ coincide with that in above
discussions.

\subsection{MSSM$\pm R_{\rm p}$}

Now, we try to apply the obtained formalism to models beyond the SM, the
MSSM$\pm R_{\rm p}$, to see whether there would be some obviously
different behaviors. For convenience to see the new contributions,
i.e. the deviations from $\Delta_j^{W_{\rm SM}}$ in the SM, we define
\begin{equation}
  \Delta_j \equiv \Delta_j^{W_{\rm SM}} \left( 1 + \delta_j \right) \; ,
  \label{eq:deltanp}
\end{equation}
where,
\begin{equation}
  \delta_j = 2 \, {\rm Re} \left[ 
  \frac{\delta G_-}{G_-^{W_{\rm SM}}} \right] 
  + \sum_{i=\pm} \left| 
  \frac{\delta G_i}{G_-^{W_{\rm SM}}} \right|^2
  + \frac{a_j \mch \, \muh^2}{\fl_j} \, {\rm Re} \left[ 
    \left( 1 + 
  \frac{\delta G_-}{G_-^{W_{\rm SM}}} \right) \, 
  \frac{\delta G_+}{G_-^{W_{\rm SM}}} \right] \; .
  \label{eq:deltas}
\end{equation}
Here in the expression, $\delta G_\pm$ contain contributions from
$\b-$mediated diagrams except for $\b = W_{\rm SM}$. 

In the MSSM\prp \cite{martin}, new contributions at tree-level are induced
only by charged Higgs bosons through the effective Lagrangian
\begin{equation}
  \l_{H^\pm} = \frac{g}{\sqrt{2} m_W} \, V_{ij} 
  \left[ \ub_i \, \left( \cot\beta \, m_{u_i} \, P_- + 
      \tan\beta \, m_{d_j} \, P_+ \right) \, d_j \right] \, H^+ 
      + {\rm h.c.} \; ,
  \label{eq:lsusyh}
\end{equation}
where $\tan\beta \equiv v_2/v_1$ is the ratio of the vacuum expectation
values of two Higgs doublets. Since the Higgs bosons couple to quarks
via the scalar interaction, after the Fierz transformation, although
in principle there are many kinds of operators, only the mixing terms,
$P_\pm P_\mp$, would exist for the Upsion production, which consequently
leads to a cancellation of the $\tan\beta$ in couplings, like
\begin{eqnarray}
  \delta G_+ = 
  \frac{1}{8} \, \frac{\mch \, \muh}{\mwh^2} \f_{H^\pm} & , &
  \delta G_- = 
  \frac{1}{8} \, \frac{\muh}{\mwh^2} \f_{H^\pm}  \; .
  \label{eq:fmch} 
\end{eqnarray}
Although the mixing terms are suppressed relative to the squared terms, 
$P_\pm^2$, by a factor of ${\muh}/{\mwh^2}$, the nature of independence of
the $\tan\beta$ is of an advantage of it, which may lead to a precise
measurement of the charged Higgs mass in the future experiments. 
Explicitly, the $\delta_j$ reads as
\begin{eqnarray}
  \delta_j & = &
  {\rm Re} \left[ \frac{1}{4} \frac{\muh}{\mwh^2}
     \frac{\f_{H^\pm}}{\f_{W_{\rm SM}}} \right]
  \left( 1 + \frac{a_j \mch^2 \muh^2}{2 \, \fl_j} \right) 
  \nonumber \\
  & & 
  + \frac{1}{4} \left| \frac{1}{4} \frac{\muh}{\mwh^2} 
    \frac{\f_{H^\pm}}{\f_{W_{\rm SM}}} \right|^2
  \left( 1 + \mch^2 + 
    \frac{a_j \mch^2 \muh^2}{\fl_j} \right) \; ,
  \label{eq:deltamssmpr} 
\end{eqnarray}
which behaves like an expansion of
\begin{eqnarray}
       \left| \frac{1}{4} \frac{\muh}{\mwh^2} \, 
        \frac{\f_{H^\pm}}{\f_{W_{\rm SM}}} \right| 
      \sim 7\% \times 
       \left| \frac{\f_{H^\pm}}{\f_{W_{\rm SM}}} \right| \; ,
       \label{e35}
\end{eqnarray}
which tends to be smaller with the increase of the charged Higgs mass
in the region outside the threshold as shown in Fig. \ref{fig:fb}.
Furthermore, the coefficients contain small corrections of the order
$\sim O(\mch^2)$ in Eq. (\ref{e35}) coming from the right-handed coupling.

As for MSSM\mrp \cite{mssmwrp}, in addition to the charged Higgs 
interactions there are also contributions due to the sfermion, squark
or slepton, mediated processes via the interactions
\begin{eqnarray}
        \l_\tl & = & \lpijk \, 
                {\tl_-}^i \, {\bar{d}_+}^k \, {u_-}^j + {\rm h.c.} \; ,
        \label{eqn:lglp}\\
        \l_\td & = & -\lppijk \, 
                {\td_+}^k \, \left( {\bar{u}_-}^i \right)^c \, {d_-}^j 
                + {\rm h.c.}  \; ,
        \label{eqn:mssmmr}
\end{eqnarray}
which are obtained by expanding the additional term in the superpotential
without $R-$parity. Here the notations of $f_\pm = P_\pm \, f$ are used
for left and right-handed particles. It should be stressed that while
both interactions in above can lead to the desired signal, they cannot
both occur simultaneously in nature since they would lead to fast proton
decay.

The sfermions give a contribution like
\begin{eqnarray}
  \delta G_+ = 0 & , &  
  \delta G_- = 
  \frac{1}{2 \sqrt{2} \gf \mwh^2} \, \left( 
  \frac{\lpp_{2i3}^\ast \lpp_{3i3}}{V_{cb}^\ast V_{tb}} \f_\td 
  + \frac{\lp_{i23}^\ast \lp_{i33}}{V_{cb}^\ast V_{tb}} \f_\tl
  \right) \; , 
  \label{eq:fml} 
\end{eqnarray}
where $i$ is the generation index with $i = 1,2(1,2,3)$ for 
$\td_i(\tl_i)$. For simplicity, here we consider only the lightest 
sfermions, i.e. either a down squark or slepton, in the calculation.
With some algebraic manipulations we get the sfermion contribution as
\begin{eqnarray}
  \delta_j^\tf & = &
  \delta_j^{{\rm MSSM}-R_{\rm p}} - 
  \delta_j^{{\rm MSSM}+R_{\rm p}}
  \nonumber \\
  & = & 
  {\rm Re} \left[ \left( \frac{r_\tf}{\sqrt{2} \gf \mwh^2} 
  \frac{\f_\tf}{\f_{W_{\rm SM}}} \right)
  \left[ 1 + 
    \frac{1}{2} 
    \left( \frac{1}{4} \frac{\muh}{\mwh^2} 
      \frac{\f_{H^\pm}}{\f_{W_{\rm SM}}} \right)
  \left( 1 + \frac{a_j \mch \muh^2}{2 \, \fl_j} \right) \right] 
  \right]
  \nonumber \\
  & & 
  + \frac{1}{4} \left| \frac{r_\tf}{\sqrt{2} \gf \mwh^2} 
    \frac{\f_\tf}{\f_{W_{\rm SM}}} \right|^2 \; ,
  \label{eq:deltamssmmr}
\end{eqnarray}
where a shorthand notation $r_\tf \equiv {\la^\ast \la}/{(V_{cb}^\ast
V_{tb})}$ is taken.

\section{Numerical Results}

\begin{figure}[t]
  \centering
    \begin{minipage}[c]{0.4\textwidth}
      \centering \includegraphics[scale=0.35]{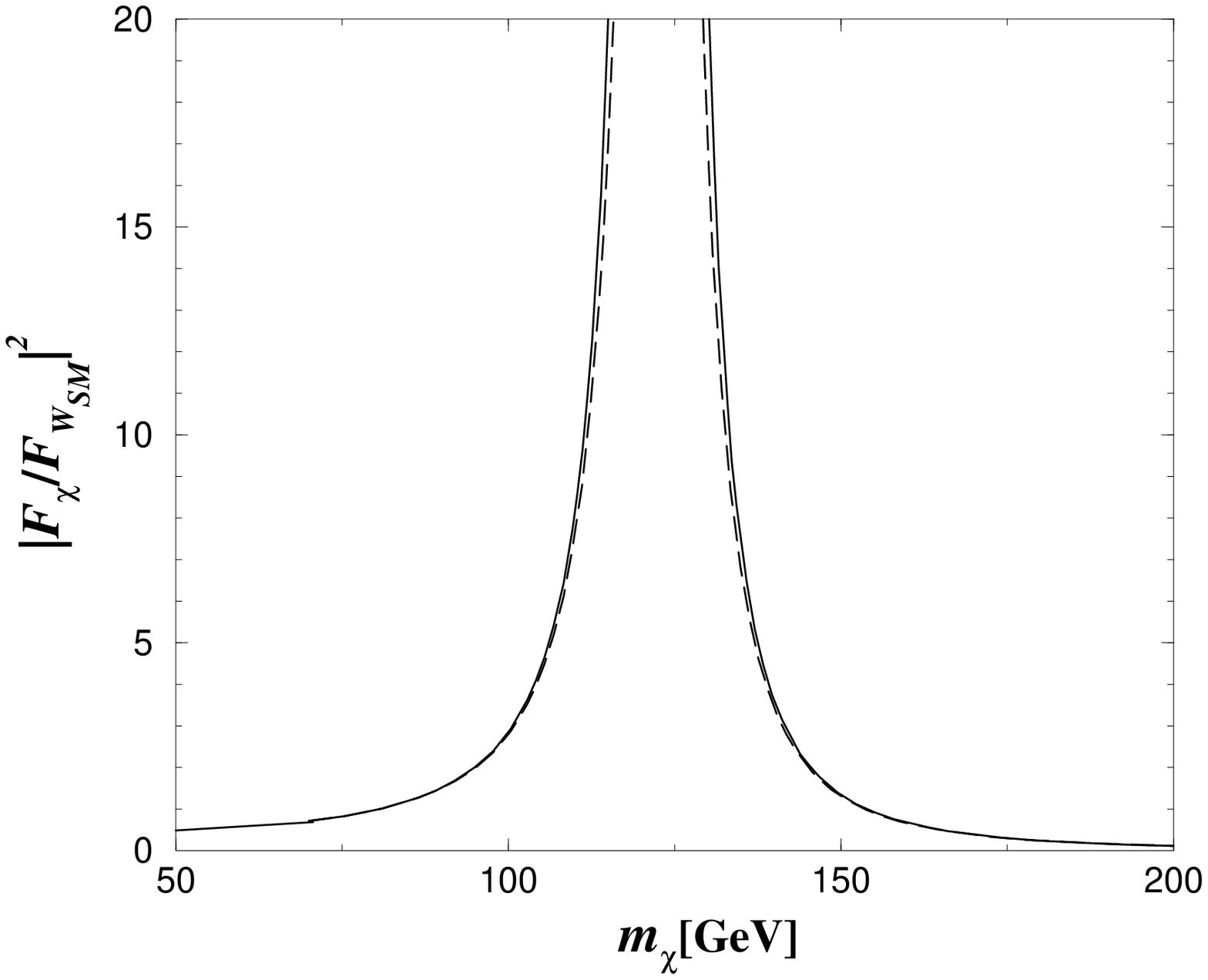}
    \end{minipage} 
    \hspace*{5mm}
    \begin{minipage}[c]{0.4\textwidth}
      \centering \includegraphics[scale=0.35]{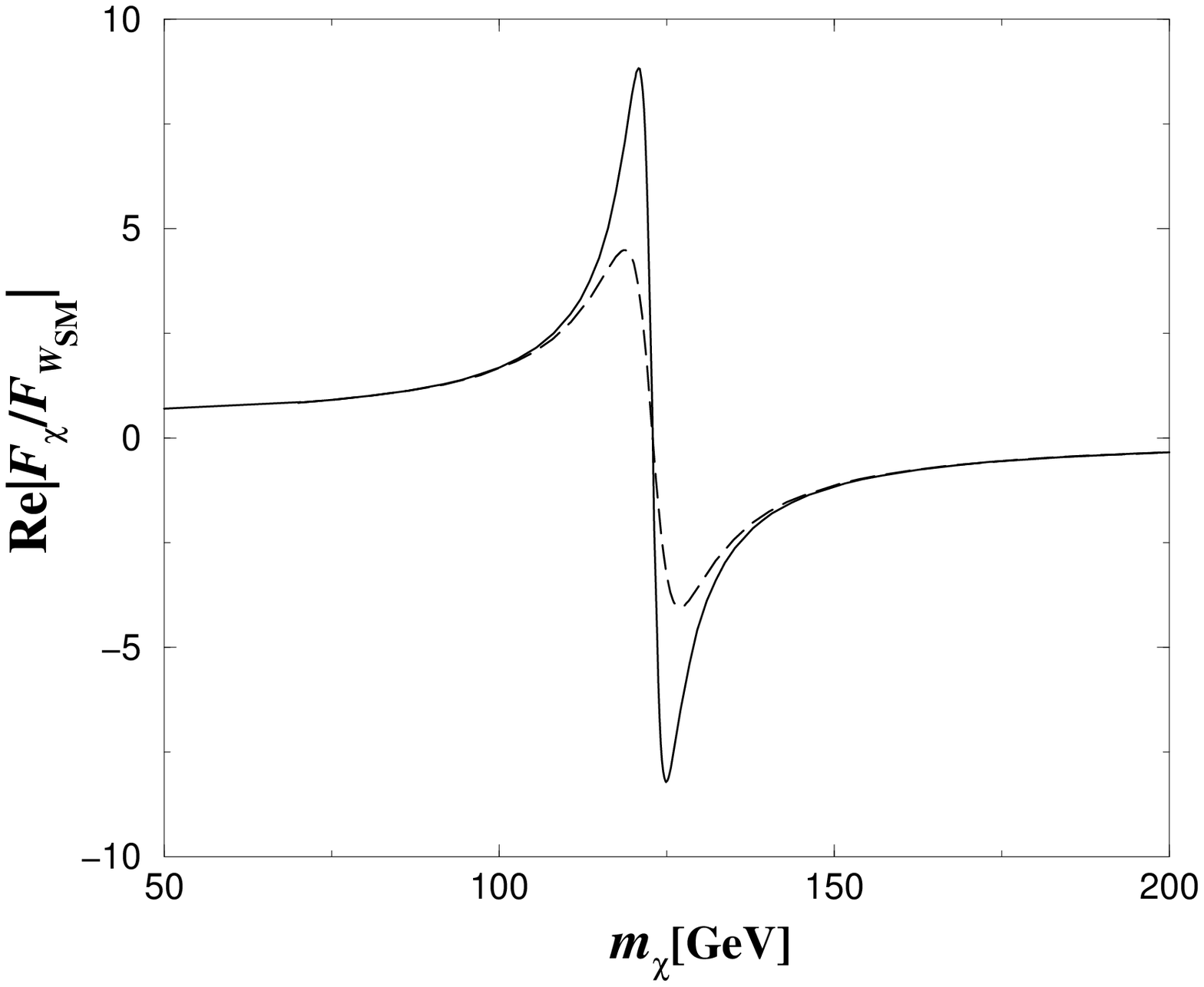}
    \end{minipage}
\caption{$\left| {\f_\b}/{\f_{W_{\rm SM}}} \right|^2$ (left) and ${\rm Re}
\left[ {\f_\b}/{\f_{W_{\rm SM}}} \right]$ (right) versus $m_\b$ with
$\gl_\b = 2 \gl_{W_ {\rm SM}}$ (solid line) and $\gl_\b = 4 \gl_ {W_{\rm
SM}}$ (dashed line).}
\label{fig:fb}
\end{figure}

In performing the numerical calculation, we take the central values of
the known parameters \cite{pdg} 
\begin{center}
   $m_c = 1.25 \pm 0.15$ (GeV) , 
   $m_b = 4.1 \sim 4.4$ (GeV) ,
   $m_t = 173.8 \pm 5.2$ (GeV) , \\
   $m_W = 80.41 \pm 0.10$ (GeV) , 
   $\gl_W = 2.06 \pm 0.06$ (GeV) , 
   $G_F = 1.166 \times {10}^{-5}$ (GeV)$^{-2}$ , \\
   $\left| V_{cb} \right| = 0.0395 \pm 0.0017$ ,
   $\sin^2\theta_W = 0.23$ . 
\end{center}
For $\b$, we presume that $m_\b$ sits in a region to be the order of
magnitude of the top quark mass, and assume its decay width to be fixed to
the order of $\gl_{W_{\rm SM}}$ in this mass region as well. Generally
speaking the decay width would depend on the mass, but here for the narrow
mass region in consideration, the fixed decay width assumption might be 
reasonable to certain degree. As shown in Figures \ref{fig:afb} and
\ref{fig:fb}, we see that the effect of varying decay width manifests
significantly only in the region around threshold, i.e. $m_\b \approx
110\sim135$(GeV).

Besides, the ratio of couplings is defined as
\begin{equation}
  r_\b \equiv \left| r_\b \right| \, {\rm e}^{i \theta_\b} \; ,
   \label{eq:rc} 
\end{equation}
since, in general, the $r_\b$ could be a complex number. The angle
$\theta_\b$ here could be an unremovable one relative to the phases of the
$6\times 6$ CKM matrix elements, which may influence the final result 
somewhat. In the case of MSSM\mrp, exploiting the bounds on $\lp$ and
$\lpp$ as given in Ref. \cite{dreiner}, we have
\begin{eqnarray}
\label{bound1}
  \left| r_{\td_i} \right| & < & 13.61 \; \; \; \; 
         {\rm for} \; \; i= 1,2 \; , \\
  \left| r_{\tl_i} \right| & < & \left\{ 
    \begin{array}{lcl}
\label{bound2}
      0.06 & {\rm for} & i = 1 \\
      1.78 & {\rm for} & i = 2 \\
      1.32 & {\rm for} & i = 3 
    \end{array}
    \right. \; .
\end{eqnarray}
Since the lightest sfermion belongs to the third generation ($i = 3$),
which gives the largest contributions in the calculation, we
will take its value, $\left| r_\tf \right| = 1.32$, in the whole
sfermion mass region in consideration. To be noted that the boundarys of
(\ref{bound1}) and (\ref{bound2}) are obtained as taking $m_\tf =
100$ GeV and they may tend to be looser with the increase of $ m_\tf$.

\begin{figure}[t]
    \centering
    \includegraphics[scale=0.6]{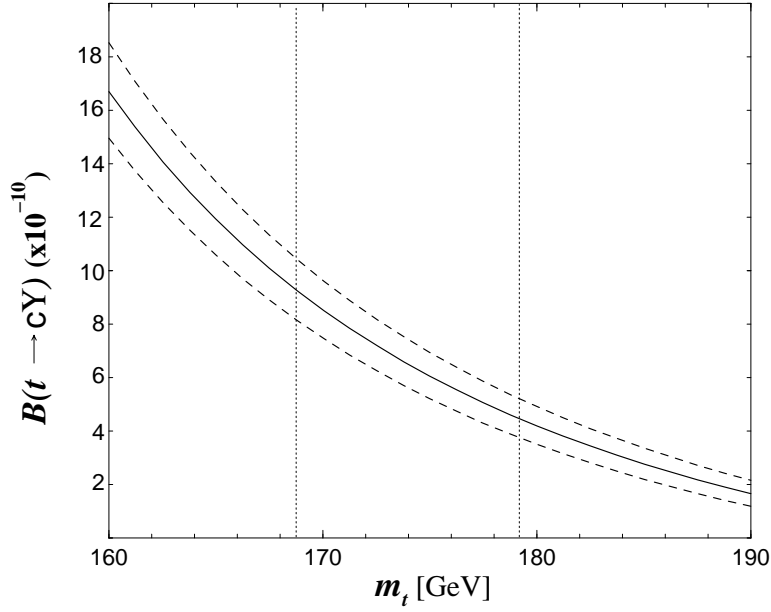}
\caption{$\br(\tcups)$ in the SM as a function of $m_t$ with the central
value $|V_{cb}|$ (solid line) and the error bar coming from $|V_{cb}|$ 
(short-dashed lines). The vertical dotted lines indicate the experimental
bounds of $m_t$.}
\label{fig:sm}
\end{figure}

With the discussions in above on the inputs, we can now proceed the
numerical evaluations. Within the SM the calculation of the branching
ratio is quite straightforward, it is
\begin{eqnarray}
\label{branch}
  \br_{\rm SM}(\tcups) & = & 
   \left( \frac{(f_\ups^1)^2}{9} 
    + 8 \sum_{i} (f_\ups^{8, i})^2 \right) \, 
  ( 1.6 \pm 0.3 \pm 0.1 ) \times {10}^{-8} (GeV)^-2
  \nonumber \\
  & = & \left( 6.4 \pm 1.2 \pm 0.5 \right) 
    \times {10}^{-10} \; .
\end{eqnarray}
Here in the evaluation, the differences between Eq. (\ref{eq:fl}) and
(\ref{eq:fl2}), which are always suppressed by $\mch \muh$ and
less than $0.5\%$, as well among $a_j$s, are all neglected; 
the $i$ in the sum runs through the various color-octet states. 
In Eq. (\ref{branch}) the errors are systematic uncertainties in 
$m_t$ and $|V_{cb}|$. These errors may induce about
ten percent of uncertainties in the result as illustrated in 
Fig. \ref{fig:sm}. The uncertainties coming from the values of 
$m_c$, $m_\ups$, $m_W$ and $f_\ups^{1, 8}$ are apparently negligible.
It is also noted that in the prompt $\ups$ production about $30\%$ of all
may come from the feeddown of its higher excited states. If experiments do
not separate these different sources, the present branching ratio should
be enhanced by the same amount.

As for the models beyond the SM, we first show the relations between
the values of ${\rm Re} \left[{\f_\b}/{\f_{W_{\rm SM}}} \right]$ and
$\left| {\f_\b}/{\f_{W_{\rm SM}}} \right|^2$ versus $m_\b$ with different
$\Gamma_\b$ in Fig. \ref{fig:fb}. Since the $m_\b$ dependence
exits only in these ratios, in fact Fig. \ref{fig:fb} gives a universal
feature of all $\b$-mediated processes. In addition, the sizes of these
two ratios are comparable and would decrease as $m_\b$ increases, except
for the region near the threshold. As a consequence of these facts, as
long as expressions contain these ratios, the contributions are dominantly
coming from terms including ${\rm Re}\left[{\f_\b}/{\f_{W_{\rm SM}}}
\right]$ in MSSM\prp \, since the prefactors involving the ratios of
couplings and /or masses are less than unity. However, for MSSM\mrp \, the
term including $\left| {\f_\b}/{\f_{W_{\rm SM}}} \right|^2$ might be
dominant supposing the factors ${r_\tf}/{\sqrt{2} \gf \mwh^2}$ are much
larger than unity.

The value of $\delta_j$, the new contribution from beyond the SM, in the
MSSM\prp \, is shown in Fig. \ref{fig:susyr} versus $m_t$ for various
$m_{H^\pm}$. For MSSM\mrp, the additional sfermion contributions to
$\delta_j^\tf$ is displayed in Fig. \ref{fig:susywr} as a function of
$m_\tf$ for various $\theta_\tf$. Here the charged Higgs mass dependence
of $\delta_j^\tf$ is quite weak, which is easy to figure out from Eq.
(\ref{eq:deltamssmmr}).

\section{Conclusions}

The top quark decay process $\tcups$ is investigated within and beyond the
SM. In dealing with the heavy quarkonium production, in this paper both
color-singlet and -octet mechanisms are employed. As a result, the decay
rates are about three orders lower than the previous rough estimations
\cite{lth} in the SM and MSSM\prp. It is found that the decay rate even 
within the SM is still much larger than that of the rare top decays.

\begin{figure}[t]
    \centering
    \includegraphics[scale=0.45]{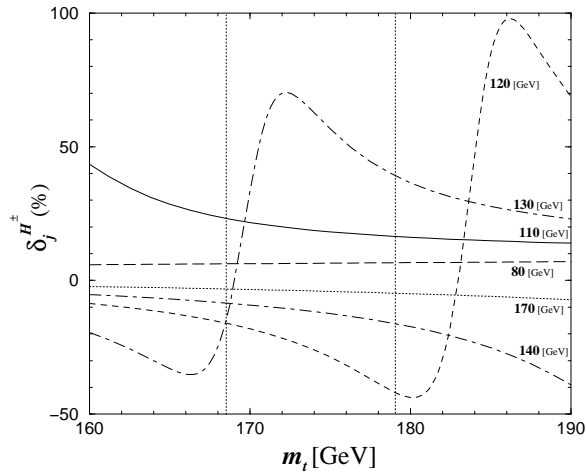}
\caption{$\delta_j^{H^\pm}$ varys as a function of $m_t$ for $m_{H^\pm}$ 
to be equal to 80, 110, 120, 130, 140, and 170 GeV, as shown in the right
rectangular, respectively. The vertical dotted lines represent the 
experimental bounds of $m_t$.}
\label{fig:susyr}
\end{figure}

\begin{figure}[t]
  \centering
  \includegraphics[scale=0.45]{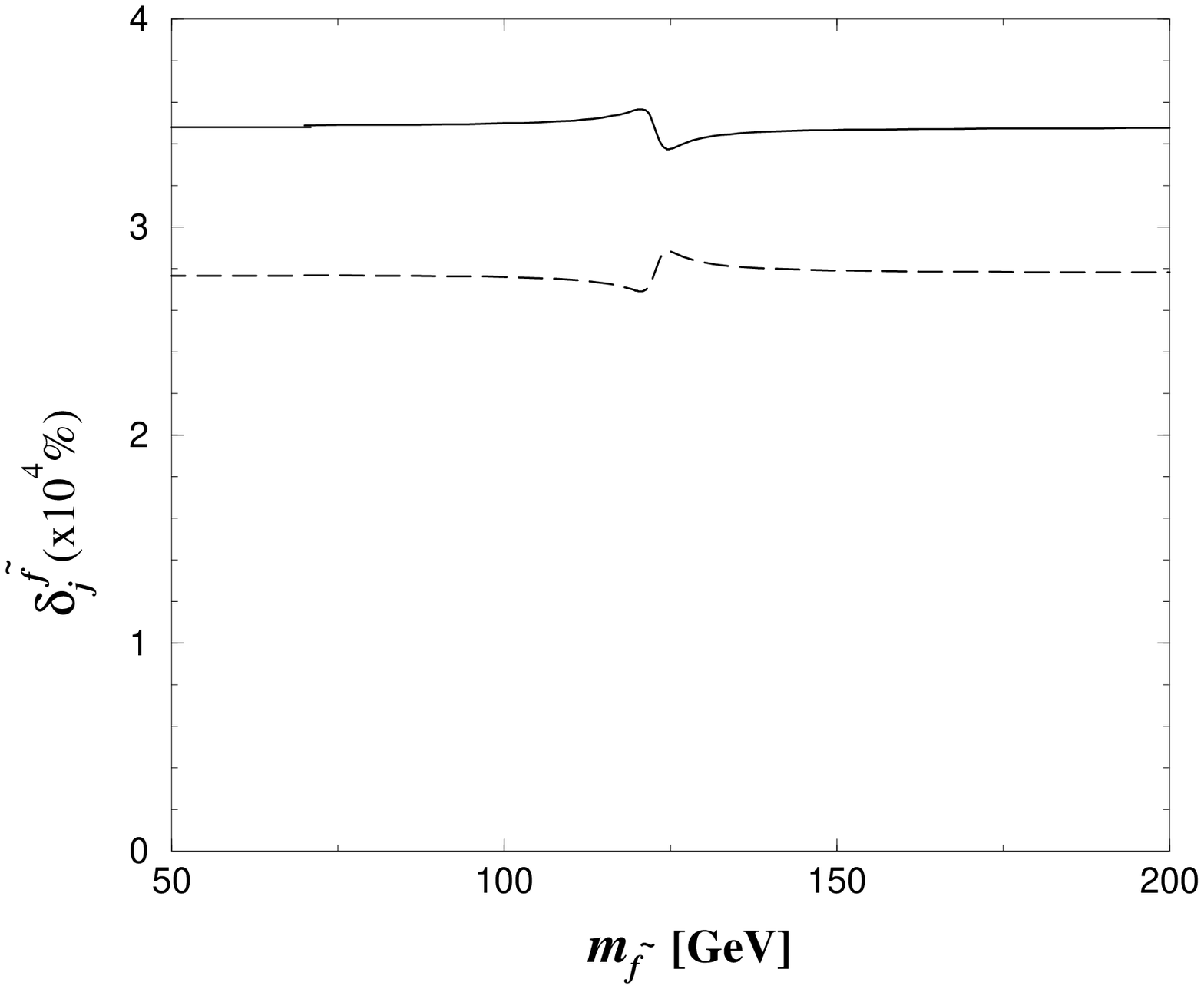}
    \caption{$\delta_j^\tf$ as a function of 
      $m_\tf$ in ${|r_\tf| = 1.32}$ with 
      $\theta_\tf = 0$ (solid line) and
      $\theta_\tf = \pi$ (dashed line).}
    \label{fig:susywr}
\end{figure}

As shown in Figs. \ref{fig:susyr} and \ref{fig:susywr}, new interactions
in models beyond the SM could enhance greatly the branching ratio.
According to Eqs. (\ref{eq:deltanp}), (\ref{eq:deltamssmpr}) and
(\ref{eq:deltamssmmr}), the branching ratio would be changed by a factor
of $\left(1 + \delta_j \right)$, i.e. $\br(\tcups) = \left(1 + \delta_j
\right) \, \br_{\rm SM}(\tcups)$, with
\begin{equation}
  \delta_j \leq \left\{ 
  \begin{array}{lcl}
  70\% & : & H^\pm \\
  35000\% & : & \tf
  \end{array}
  \right. \; 
\end{equation}
in the presumed mass region given in Figs. \ref{fig:susyr} and
\ref{fig:susywr}, where slightly light charged Higgs masses are taken
schematically accounting for the fact that light charged Higgs mass, less
than 100 GeV, is still not excluded by experiments \cite{pdg}. The second
enormous supplement of $\delta_j$ stems from the enhancement of prefactor 
${r_\tf}/{\sqrt{2} \gf \mwh^2}$ in (\ref{eq:deltamssmmr}). Because
of the fact that $|r_\tf|$ only runs slightly with the $m_\tf$ 
within the considered region, it should be pointed out that in drawing
Fig. \ref{fig:susywr} we merely treat it fixed.

Since in the next round of experiment, such as LHC, there are about
${10}^6 \sim {10}^8$ top quark events would be generated per year, it is
obvious that within the SM the present discussed process is still at the
unreachable level. Therefore, we hope this decay process may
play a role in probing new physics in the near future. 
Unlike the cases of Quarkonium production at B-meson and Z-boson decays,
here it is found that the color-octet contributions are negligable,
although the relative coefficient enhancement still existing. This nature
makes our calculations on a more reliable base, since it is well-known
that large uncertainties still remain in the values of color-octet matrix
elements among different fittings. To make a more precise estimation, 
the more accurate values on $m_t$ and $|V_{cb}|$ are required. Last, 
the calculation shows that this top quark decay process provides a 
possibility of measuring the charged Higgs mass without the $\tan\beta$ 
dependence, as shown in Eqs. (\ref{eq:fmch}) and
(\ref{eq:deltamssmpr}). 
\vspace{1mm}

\centerline{\large \bf Acknowledgments} 

\noindent
LTH thanks High Energy Division of the Abdus Salam ICTP for the warm
hospitality while part of this work was done. The authors acknowledge
Ahmed Ali for reading the manuscript and some useful comments.

\clearpage

\begin{thebibliography}{99}
\bibitem{r1} CDF Collaboration, F.Abe {\it et al.}, Phys. Rev. Lett.  
{\bf 74}, (1995) 2626.

\bibitem{r2} CDF Collaboration, F.Abe {\it et al.}, Phys. Rev. {\bf D50}, 
(1994) 2966.

\bibitem{r3} D0 Collaboration, S.Abachi {\it et al.}, Phys. Rev. Lett. 
{\bf 74}, (1995) 2632.

\bibitem{2hdm} G. Eilam, J. L. Hewett and A. Soni, 
\journal{\pr} {D44} {1991} {1473}.

\bibitem{mele} B. Mele, S. Petrarca and A. Soddu, 
\journal{\pl} {B435} {1998} {401}.

\bibitem{gunion} B. Grzadkowski, J. F. Gunion and P. Krawczyk,
\journal{\pl}{B268} {1991} {106}.
        
\bibitem{susy} C. S. Li, R. J. Oakes and J. M. Yang, \journal{\pr} {D49}
{1994} {293}; 
J. M. Yang and C. S. Li, \journal{\pr} {D49} {1994} {3412};
G. Couture, C. Hamzaoui and H. K$\ddot{\rm o}$nig, \journal{\pr} {D52}
{1995} {171}; 
J. L. Lopez, D. V. Nanopoulos and R. Rangarajan, \journal{\pr} {D56}
{1997} {3100}; 
G. Couture, M. Frank and H. K$\ddot{\rm o}$nig, \journal{\pr} {D56} {1997}
{4213}; 
G. M. de Divitiis, R. Petronzio and L. Silvestrini, \journal{\np} {B504}
{1997} {45}; J. M. Yang, B.-L. Young and X. Zhang, \journal{\pr}{D58}
{1998} {055001}.

\bibitem{fermilab} {\it Future Electroweak Physics at the Fermilab
Tevatron}, report of the TeV2000 working group, \journal{FERMILAB-PUB
96/082}, {1996}, edited by D. Amidei and R. Brock.
        
\bibitem{lth} L. T. Handoko, \journal{\nc}{A111}{1998}{1275}; L. T.
Handoko and J. Hashida, \journal{\pr} {D58} {1998} {094008}.
        
\bibitem{qcd} E. Braaten and T. C. Yuan, \journal{\prl} {71} {1993}
{1673}; 
G. T. Bodwin, E. Braaten and G. P. Lepage, \journal{\pr} {D51} {1995}
{1125}; 
Err. \journal{ibid.} {D55} {1997} {5853}; 
E. Braaten and T. C. Yuan, \journal{\pr} {D52} {1995} {6627}.
        
\bibitem{qiao} F. Yuan, C.-F. Qiao and K. T. Chao, \journal{\pr} {D57}
{1998} {610}.

\bibitem{surplus} For example: E. Braaten and S. Fleming, Phys. Rev. Lett.
{\bf 74} (1995) 3327; 
K. Cheung, W.Y. Keung, and T.C. Yuan, Phys. Rev. Lett. {\bf 76} (1996)
877; 
C.-F. Qiao, F. Yuan, and K.T. Chao, Phys. Rev. {\bf D55} (1997) 4001.

\bibitem{schuler} Gerhard A. Schuler, report-no. CERN-TH-7170-94,
hep-ph/9403387; and references therein.

\bibitem{pdg} C. Caso {\it et.al.} (PDG Collaboration), \journal{\epj}
{C3} {1998} {1}.

\bibitem{qcd8} P. Cho, A. K. Leibovich, \journal{\pr} {D53} {1996} {150}; 
\journal{ibid.} {D53} {1996} {6203}.
        
\bibitem{tbw} I. Bigi, Y. Dokshitzer, V. Khoze, J. K$\ddot{\rm u}$hn and
P. Zerwas, \journal{\pl} {B181} {1986} {157}; 
M. Je$\dot{\rm z}$abek and J.H. K$\ddot{\rm u}$hn, \journal{\np} {B320}
{1989} {20}.
        
\bibitem{martin} See for example, J. F. Gunion and H. E. Haber, 
\journal{\np} {B272} {1986} {1}; \journal{ibid.} {B278} {1986} {449}.
        
\bibitem{mssmwrp} See for example, C. S. Aulakh and R. N. Mohapatra, 
\journal{\pl} {B119} {1982} {316}.
        
\bibitem{dreiner} H. Dreiner, {\it Perspectives on Supersymmetry}, 
edited by G. L. Kane, World Scientific, Singapore, 1999, 462-479.

\end{thebibliography}
\end{document}